\documentclass{elsart}
\usepackage{graphicx}
\usepackage{amsmath}
\usepackage{amssymb}
\usepackage{epsfig}
\usepackage{dcolumn}
\usepackage{bm}
\begin{document}
\begin{frontmatter}

\title{Self-organization of quasi-equilibrium stationary condensation in accumulative
ion-plasma devices}
\author{V.I. Perekrestov $^{a}$, A.I. Olemskoi $^{a,b}$, Yu.A. Kosminska $^{a}$,}
\author{A.A. Mokrenko $^{a}$}
\address{$^{a}$ Sumy State University, 2, Rimskii-Korsakov St., 40007 Sumy,
Ukraine}
\address{$^{b}$ Institute of Applied Physics, 58, Petropavlovskaya St., 40030 Sumy,
Ukraine}

\date{}

\begin{abstract}
We consider both theoretically and experimentally self-organization process of
quasi-equilibrium steady-state condensation of sputtered substance in
accumulative ion-plasma devices. The self-or\-ga\-ni\-za\-tion effect is shown
to be caused by self-consistent variations of the condensate temperature and
the supersaturation of depositing atoms. On the basis of the phase-plane
method, we find two different types of the self-organization process to be
possible. Experimental data related to aluminum condensates are discussed to
confirm self-organization nature of quasi-equilibrium steady-state condensation
process.
\end{abstract}

\end{frontmatter} \maketitle

\section{Introduction}\label{Sect.1}

Formation of various nanosystems in the course of the condensation process
is known to be achieved due to widespread technologies such as the
molecular-beam epitaxy, the metal-organic vapour-phase epitaxy, the
liquid-phase epitaxy, the electrolytic deposition etc. \cite{1,2}.
Characteristic peculiarities of these technologies are both quasi-equilibrium
and steady-state to be as follows:
 \begin{itemize}
\item the condensation process occurs at sufficiently low
supersaturation to provide a proximity to phase equilibrium between condensate
and depositing substance;
\item this supersaturation remains stable in the course of the time
to guarantee steady-state conditions of the condensation process.
 \end{itemize}
Due to above conditions adsorbed atoms can be coupled with growing condensate
surface only if the strongest chemical bonds are realized to minimize the free
energy \cite{bib:A,bib:C}.

The proximity to the phase equilibrium is known to be governed by decrease of
the supersaturation
\begin{equation}
\label{1} \xi=(n-n_{\rm e})/n_{\rm e}
\end{equation}
where $n$ and $n_{\rm e}$ are current and equilibrium concentrations of atoms
above the growth surface. In the case of vapour-condensate systems, the latter
of this concentrations is described with the empirical relation
\begin{equation}
\label{2} n_{\rm e}=\frac{A(T)}{k_BT}\exp\left(-\frac{E_{d}}{k_BT}\right)
\end{equation}
where $E_{d}$ is the desorption energy, $T$ is the growth surface
temperature, $k_B$ is the Boltzmann constant; the temperature
dependent parameter $A(T)\hm=\exp(\alpha+\beta T+\gamma/T)$ is
determined with a set of constants $\alpha$, $\beta$ and $\gamma$
being characteristics for a given substance \cite{bib:D}.

In the case of \emph{volatile} substances, the desorption energy is so small
$(E_{d}\ll k_BT)$ to expand large value of the equilibrium concentration
(\ref{2}) into a series. As a result, the supersaturation becomes less
sensitive to temperature change that simplifies implementation of
quasi-equilibrium steady-state condensation at relatively high deposition
fluxes.

For the \emph{vapour-condensate} systems related to the molecular-beam epitaxy,
low supersaturations are obtained only due to increased growth surface
temperature at relatively weak deposition fluxes. That is why the
molecular-beam epitaxy can be implemented mainly for those substances which
have relatively high volatility.

In \emph{chemically active medium-condensate} systems, being the basis of
metal-organic vapour-phase epitaxy, liquid-phase epitaxy and electrolytic
deposition, the proximity to the phase equilibrium is stimulated additionally
by reversible chemical processes. The latters induce reverse transitions of
adsorbed atoms which are weakly coupled with growth surface into environment.
In Eqs. (\ref{1}) and (\ref{2}) this is expressed in decreasing value $E_{d}$
of the desorption energy down to an effective value that increases the
equilibrium concentration $n_{\rm e}$ stimulating proximity to equilibrium.
From this point of view, above chemical methods make it possible to reach low
supersaturations at condensation of weakly volatile substances that gains its
advantage over the molecular-beam epitaxy.

The present work is devoted to consideration of the system \emph{low
temperature plasma-condensate} which enables to reach a proximity to phase
equilibrium by means of heating growth surface with the help of plasma stream,
both momentum and energy transfer from plasma particles to adatoms, and their
partial thermal accommodation. Due to the latter two mechanisms the desorption
energy $E_{d}$ gets an effective value \cite{bib:A,bib:E,bib:F}
\begin{equation}
\label{3} E=E_{d}-\delta E
\end{equation}
with a stochastic addition $\delta E$ characterized by the mean value $\bar{E}$
and the dispersion $\sigma^2\equiv\overline{\left(\delta E-\bar{E}\right)^2}$
(the overline means averaging over plasma particle collisions). Thus, in
analogy with the chemical systems the low-temperature plasma-condensate systems
make it possible to stimulate proximity to equilibrium at condensation of
weakly volatile substances. However, making use of the plasma arrives at the
problem to fix steady-state conditions during the condensation process.

For all mentioned technologies, there arises a common range of problems to fix
steady-state regime of the condensation process at vanishing supersaturations.
The main point is that such crucial technological parameters as the growth
surface temperature $T$ (hence, the equilibrium concentration $n_{e}=n_{e}(T)$)
and the depositing flux $J$ are mutually independent and are regulated by
different power sources. Also, under quasi-equilibrium conditions the
condensing flux $J_{c}$ is obeyed the inequality $J_{c}\ll J$. Hence, even
slight relative variations $\Delta J/J$ of the depositing flux can arrive at
considerable changes in condensation kinetics at condition $\Delta J \sim
J_{c}$. (The same effect can cause variation of the equilibrium concentration
$n_{e}(T)$ at slight fluctuations of the growth surface temperature $T$.) In
the absence of self-organization, these problems are solved by artificially
created feedback between condensation kinetics and depositing flux by means of
control systems that arrives at considerable rise in the cost of technologies.

Therefore, the aim of this Letter is two-fold: i) elaborate a
technological device, whose operation within self-organization
regime ensures extremely low steady-state supersaturations; and ii)
build up a theoretical model to describe above self-organized
process. In correspondence with these aims, the outline of the paper
is as follows. In Section \ref{Sect.2}, we state physical
backgrounds of operation of the accumulative ion-plasma devices. On
the basis of these conceptions, we consider theoretically the
self-organization of extremely low stationary supersaturations in
Section \ref{Sect.3}. Experimental validation of the
self-organization scheme is given in Section \ref{Sect.4}, and
Section \ref{Sect.5} concludes our consideration.

\section{Physical backgrounds of operation of the accumulative ion-plasma
device}\label{Sect.2}

\begin{figure}[!h]
\centering
\includegraphics[width=130mm]{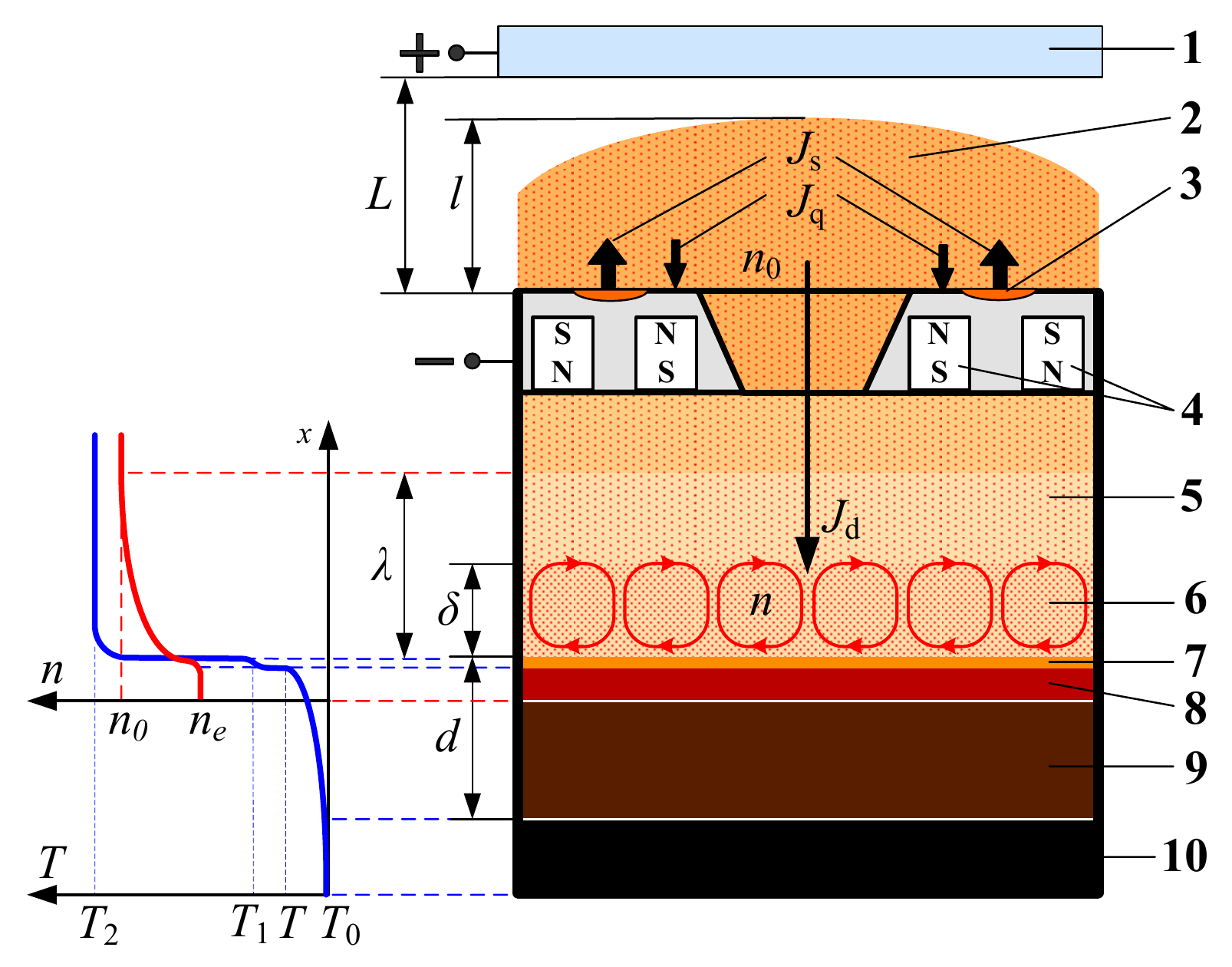}
\caption{ (Colour online) Scheme of axisymmetric ion-plasma device (1 -- anode,
2 -- thermalization volume of sputtered atoms, 3 -- erosion zone, 4 -- magnet
system, 5 -- hollow cathode, 6 -- region of circular mass transfer, 7 -- layer
of adsorbed atoms, 8 -- condensate, 9 -- substrate, 10 -- chiller); plot on the
left depicts both temperature and concentration distributions near growth
surface 7.} \label{Fig_1}
\end{figure}

Principle components of our sputtering device are the magnetron
sputterer in combination with the hollow cathode which operate under
high pressure of highly refined argon ($P_{Ar}=10\div 30 Pa$) (see
Fig.\ref{Fig_1}). Such a construction
allows to stabilize and increase the discharge current due to both hollow
cathode and the magnetron effects \cite{bib:G,bib:H}, the last of which
localizes the erosion zone in addition. It is worthwhile to stress that almost
all atoms of inert gas and sputtered substance become ionized when get inside
the hollow cathode \cite{bib:G,bib:H}. Under increased pressure, plasma
particles collide so frequently and intensively to average their energy
effectively. These collisions decrease the dispersion $\sigma$ of the effective
desorption energy (\ref{3}) that, in turn, stabilizes condensation process.

The self-organization process starts with accumulation of a sputtered substance
near the growth surface. According to Ref. \cite{bib:I}, atoms sputtered by
magnetron lose their energy (in other word, thermalize) due to collisions of
particles of high-pressure plasma, so that their further movement becomes
diffusive. Related thermalization length is estimated as
$l\propto(T_{Ar}+\Delta T)/T_{Ar}P_{Ar}$ with $T_{Ar}$ and $\Delta T$ being the
temperature of unheated working gas and its growth under discharge. Hence, the
thermalization length is reduced with increasing the pressure $P_{Ar}$ and
decreasing discharge power that reduces the temperature growth $\Delta T$. As
show estimations \cite{bib:H,bib:I}, the flux of non-thermalized atoms in
direction of the anode can be considered as negligible at the pressure
$P_{Ar}\hm=20\div 40$ Pa and the discharge power less than 100 W, when the
thermalization length is subject to the condition $l\leq L/4$ with respect to
the target-anode distance $L$ depicted in Fig.\ref{Fig_1}.

As it is seen from this figure, the flux $J_{s}$ of sputtered substance
condenses mainly on the surface adjacent to the erosion zone with the flux
$J_{q}$ and diffuses partially into the hollow cathode with the flux $J_{d}$.
(It should be noted that at the initial time of the device operation, there is
only inert gas plasma inside the hollow cathode.) With power up, the
concentration of sputtered atoms $n_{0}$ at the hollow cathode entry is
determined by the fluxes $J_{e}$ and $J_{P}$, the first one being drift flux
due to the presence of electrical field near the inlet, and the second one
being caused by pressure difference between interior and entry of the hollow
cathode. As a result, variation of the atom concentration $n_{0}$ at the hollow
cathode inlet is governed by the equation
\begin{equation}
\label{4} \dot{n}_{0}\propto(J_{s}-J_{d}-J_{q})-(J_{e}-J_P)
\end{equation}
where and hereinafter the point above a symbol denotes the time derivative. As
the hollow cathode volume is rather small, the drift components $J_{e}$ and
$J_{P}$ become equal quickly and then sputtered substance penetrates into the
hollow cathode due to diffusion only. Because both components $J_{s}$ and
$J_{q}$ are much more than the diffusive flux $J_{d}$, one can suppose its
variation affects slightly on the concentration $n_{0}$. At the steady-state
condition $J_{d}=J_{s}-J_{q}$, diffusion process determines accumulation of
thermalized atoms in the interior of the hollow cathode to increase their
concentration over a thermodynamic threshold $n_{\rm e}$ needed to start
condensation of the sputtered substance on the substrate.\footnote{It is
worthwhile to note, under plasma influence, main part of deposited atoms are
reevaporated to be next ionized again and returned onto this surface under
electric field \cite{bib:G}. Such a circular mechanism of mass transfer
accumulates additionally depositing atoms in immediate vicinity of the growth
surface.}

To understand the lowering of steady-state supersaturation in the course of the
condensation process, let us assume the growing surface has reached such a
temperature, when inequality $n\geq n_{\rm e}$ starts to be satisfied due to
accumulation of sputtered substance. Then, the supersaturation (\ref{1}) takes
a positive value to initiate the barrier nucleation process that increases the
growth surface temperature due to the atom thermal accommodation. On the other
hand, the flux of depositing atoms decreases due to rupture of circular fluxes
that, in turn, reduces the supersaturation to minimal values providing
condensation process. In the following section, we shall show that
self-consistent variations of the surface temperature and the supersaturation
is a cornerstone of self-organization process ensuring extremely low
steady-state supersaturations.

\section{Theoretical consideration of self-organization of steady-state
quasi-equilibrium condensation}\label{Sect.3}

Let us start with consideration of main reasons for variation of the surface
temperature in the course of the ion-plasma device operation. First, each of
plasma ions transfers to the surface the average energy
$k_B\left(T_{2}-T_{1}\right)$ determined by the temperature difference of
plasma ions, $T_{2}$, and adsorbed atoms, $T_{1}$, respectively (see the
temperature distribution depicted in the left panel of Fig.\ref{Fig_1}). Then,
the energy delivered per unit time from the plasma to the unit surface can be
written in the form
\begin{equation}
\label{5} \dot{E}_{1}=\chi\theta\left(T_{2}-T\right).
\end{equation}
Here, $\chi$ is a parameter being the production of the Boltzmann
constant and the plasma flux falling down the growth surface,
$\theta\equiv(T_{2}-T_{1})/(T_{2}-T)\simeq 1-T_{1}/T_{2}={\rm
const}$ is thermal accommodation coefficient of adsorbed atoms
rewritten with accounting the condition $T\ll T_{2}$. Second, the
rate of energy transfer to the growth surface due to thermal
accommodation of condensed atoms
\begin{equation}
\label{6} \dot{E}_{2}=k_B(T_{2}-T)J_{c}
\end{equation}
is determined by the temperature difference of plasma and condensate
accompanied with the condensing flux
\begin{equation}
\label{7} J_{c}=\frac{\delta}{\tau}(n-n_{\rm e}).
\end{equation}
Here, $\delta$ and $\tau$ are characteristic length and time of the circle
motion of condensing atoms near the growth surface. Finally, the energy removal
to the chiller per unit time by means of thermal conductivity is determined by
the obvious expression
\begin{equation}
\label{8} \dot{E}_{3}\simeq\frac{\eta}{d}(T_{0}-T)
\end{equation}
where $T_{0}$ stands for the chiller temperature, $d$ and $\eta$ are the total
thickness and effective thermal conductivity of two-layer system consisting of
condensate and substrate.

Combination of the equations (\ref{5}) -- (\ref{8}) with obvious definition of
the variation rate of the surface temperature
\begin{equation}
\label{9} c\dot{T}=\dot{E}_{1}+\dot{E}_{2}+\dot{E}_{3}
\end{equation}
arrives at the differential equation
\begin{equation}
\begin{split}
\label{9a} c\dot{T}=\left[\left(\chi\theta
T_{2}+\frac{\eta}{d}T_0\right)+\frac{k_Bn_{\rm e}\delta}{\tau}T_2\xi\right]\\
-\left[\left(\chi\theta+\frac{\eta}{d}\right)+\frac{k_Bn_{\rm
e}\delta}{\tau}\xi\right]T
\end{split}
\end{equation}
with $c$ being the surface heat capacity of the growth surface. The
equation (\ref{9a}) shows the time dependence of the surface
temperature is determined by the following set of values: plasma
parameters $\chi$, $\delta$ and $\tau$; condensate-substrate
parameters $\eta$, $c$ and $d$; temperatures characterising
condensate ($T$), plasma ($T_{1}, T_{2}$, $\theta\simeq 1-T_1/T_2$)
and chiller ($T_{0}$); equilibrium concentration $n_{\rm e}$ defined
with Eq.(\ref{2}); and finally, supersaturation $\xi$ given by
Eq.(\ref{1}). Remarkably, both square brackets in Eq.(\ref{9a})
depend on the temperature $T$ only through the equilibrium
concentration (\ref{2}), whereas the last term in right-hand side
contains the factor $T$ itself. Moreover, the last terms in both
square brackets are proportional to the supersaturation $\xi$, whose
time dependence must be found next.

At finding the equation for the concentration $n$, one should take into account
that condensation occurs onto the inner surface of the hollow cathode with
large area $S$, whereas the diffusive flux penetrates toward the cathode
interior throughout the inlet hole with small area $s$. Then, the rate of the
concentration $n$ variation is determined by the difference between incoming
diffusive and outgoing condensing fluxes so that, with accounting that
accumulation of substance takes place inside the region with thickness $\delta$
near the growth surface, one obtains
\begin{equation}
\label{10} \dot{n}=\frac{1}{\delta}\left(\frac{s}{S}J_{d}-J_{c}\right).
\end{equation}
According to Eq.(\ref{10}) the steady-state regime is reached under the
condition
\begin{equation}
\label{10b} sJ_{d}=SJ_{c}
\end{equation}
which arrives at the relation $J_{c}< J_{d}$ due to the non-equality $s< S$.
The condensing flux is determined by Eq.(\ref{7}), as the diffusive component
is given by the Onsager-type expression
\begin{equation}
\label{11}J_{d}\simeq-\frac{ D }{\lambda}(n-n_{0})
\end{equation}
with effective length $\lambda$, along which the concentration varies from
$n_0$ to $n$, and the temperature dependent diffusion coefficient $ D \sim
T_{2}^{3/2}$ \cite{bib:I}. Taking into account Eqs. (\ref{2}), (\ref{7}),
(\ref{10}) and (\ref{11}), we yield the final form of the equation of motion
for the supersaturation (\ref{1}):
\begin{equation}
\label{12} \dot{\xi}+B(T)\left(1+\xi\right)\dot{T}=\frac{s}{S}\frac{ D
}{\lambda\delta}\xi_0-\left( \frac{1}{\tau}+\frac{s}{S}\frac{D
}{\lambda\delta}\right)\xi
\end{equation}
where $B(T)=\beta-\frac{1}{T}+\frac{(E_d/k_B)-\gamma}{T^2}$ and a magnitude
$\xi_0=\left(n_0-n_{\rm e}\right)/n_{\rm e}$ relates to the inlet hole of the
hollow cathode. Similarly to the differential equation (\ref{9a}), the first
term in right-hand side of Eq. (\ref{12}) depends on the temperature only
through the equilibrium concentration (\ref{2}), whereas the second term is
proportional to the supersaturation with a constant factor inside the
parentheses.

\begin{figure}[!h]
\centering
\includegraphics[width=95mm]{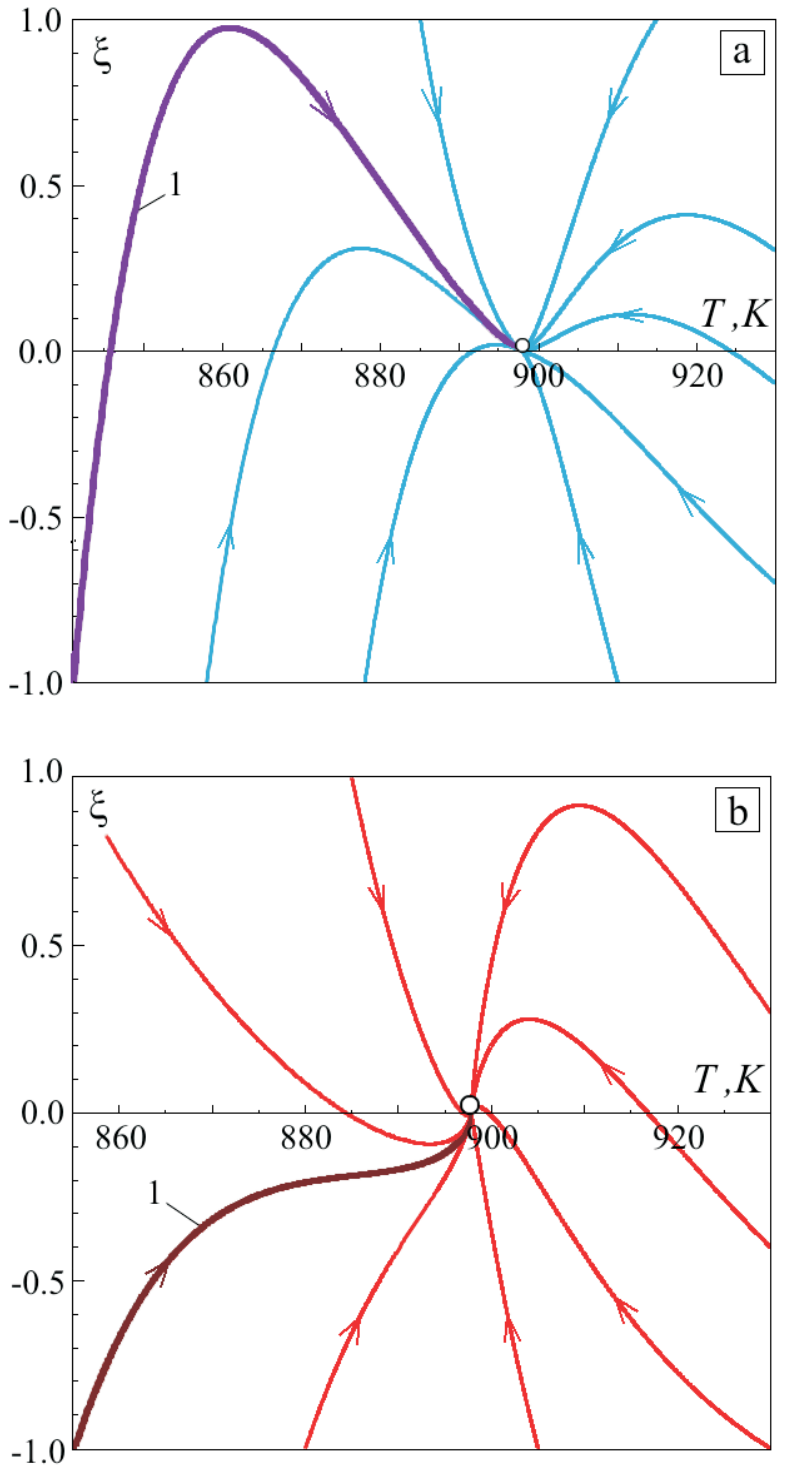}
\caption{(Colour online) Phase portraits presenting solutions of Eqs.
(\ref{9a}) and (\ref{12}) at $s/S=1/4$ (a) and $s/S=1/10$ (b) (the rest of
parameters is as follows: $D=10^5 cm^{2}s^{-1}$, $E=2.674\cdot 10^{-20} J$,
$d=0.1 cm$, $\lambda=1 cm$, $\delta=0.1 cm$, $\tau=10^{-6}s$, $n_{0}=10^{8}
cm^{-3}$, $\theta=0.95$, $\chi=7.4\cdot 10^{-3} Jìæ^{-1}cm^{-2}s^{-1}$,
$T_{2}=6000 K$, $T_{0}=300 K$, $c=3\cdot 10^{-3} Jìæ^{-1}$, $\eta=6\cdot
10^{-3} Wìcm^{-1}ìK^{-1}$ (for glass). The node corresponds to $T = 898 K$,
$\xi=1.1\cdot 10^{-4}$, $J_{c} = 1.1\cdot 10^{8} cm^{-2}s^{-1}$ (a) and $T =
898 K$, $\xi=7.4\cdot 10^{-5}$, $J_{c} = 7.4\cdot 10^{7} cm^{-2}s^{-1}$
(b).}\label{Fig_2}
\end{figure}

The system of the ordinary differential equations (\ref{9a}) and
(\ref{12}) describes self-con\-sis\-tent variations of the surface
temperature $T$ and the supersaturation $\xi$ during the
self-or\-ga\-ni\-za\-tion process of extremely low steady-state
supersaturations. The form of possible solutions of these equations
is seen from the phase portraits depicted in Fig.\ref{Fig_2}
where the set of parameters is selected applicable to the aluminum.
Analysis of the phase portraits arrives at the conclusion that
regime of the self-organization is governed specifically by the
ratio $s/S$. In the case of high values $s/S$ (Fig.\ref{Fig_2}a),
with the temperature growth the representative phase path 1 goes
through high supersaturations, whereas in the opposite case
(Fig.\ref{Fig_2}b) negative values $\xi$ precede to the steady state
related to the node. From the physical point of view, such a
behaviour is explained by that accumulation of the sputtered
substance is in advance of heating of the growth surface, in the
first case, whereas at small ratios $s/S$ slight accumulation rate
enables positive supersaturations $\xi$ only near the steady state.
From the technological point of view, the second regime is much more
preferable in case of little deposition time.

Under the steady-state conditions $\dot{T}=\dot{\xi}=0$, the system of Eqs.
(\ref{9a}) and (\ref{12}) shows that the stationary supersaturation increases
monotonically with growing effective desorption energy $E$, so that extremely
low supersaturations $\xi\ll 1$ are achieved only when $E<0.1$ eV. Usually, the
bare desorption energy for metals is estimated as $E_{d}\sim 0.4$ eV
~\cite{bib:D}, and the proximity to the phase equilibrium is provided by the
plasma influence that reduces the bare value $E_d$ to the effective energy $E$
in accordance with Eq.(\ref{3}).

\section{Experimental validation of self-organization}\label{Sect.4}

We have considered above the self-organization process of
quasi-equilibrium steady-state condensation of sputtered substance
in accumulative ion-plasma devices. At the same time, both numerical
simulations ~\cite{bib:mod1,bib:mod2,bib:mod3,bib:mod4_exp1} and
experimental investigations
\cite{bib:A,bib:mod4_exp1,bib:exp2,bib:exp3,bib:exp4,bib:K} display
a secondary self-organization process. In the course of such a
process, three-dimensional micro- and nanostructures are arranged on
the substrate to possess complicated architecture with narrow
distributions over sizes and forms of structural elements. In
contrast to the self-organization of the system plasma-condensate,
the secondary self-organization involves in the regime of the {\it
self-organized criticality} driven by fluctuations \cite{OKK}.
Really, the result of the primary process is that two phase system
plasma-condensate is staying near stationary equilibrium, but
strongly above a critical point, so that a phase transition is
thermodynamically impossible. However, appearance of active centers,
playing the role of space distributed fluctuations, initiate the
condensation process according to the following
scenario.\footnote{The condensation mechanism is beyond the scope of
the equilibrium thermodynamics to be defined by both kinetics
~\cite{bib:C} and synergetics \cite{OKK}.}

In the course of the deposition process, adsorbed substance is
embedded atom-by-atom into the growing surface on active centers
with the highest energies of chemical bonds. Among such centers, one
can point out the mono-step bends on the growing crystal surface,
inhomogeneities on the atomic-rough surface, regions of crystallite
joining and so on \cite{bib:D}. Due to the reduced surface density
of the active centers, the chemical bonds spectrum should be
appreciably discrete. That allows for one to separate those centers
whose energies of chemical bonds $E_i$, $i=1,2,\dots$ is lower than
a critical threshold $E_{c}$ fixed by external
conditions.\footnote{Energies of chemical bonds $E_i$ are considered
as their absolute values.} On these centers, the atom-by-atom
formation of the condensate becomes hardly probable because of
subcritical energies of chemical bonds $E_i<E_c$. By definition, the
threshold $E_{c}$ relates to the desorption energy $E_{d}$ in the
equilibrium concentration (\ref{2}) at zero supersaturation
(\ref{1}) and given concentration $n$ and temperature $T$. According
to Eqs. (\ref{1}) and (\ref{2}), the critical energy of chemical
bonds
\begin{equation}
\label{4-1} E_{c}=k_BT\ln\left[\frac{A(T)}{nk_{B}T}\right]
\end{equation}
increases with tending to the phase equilibrium due to concentration
decrease and temperature growth.

\begin{figure}[!h]
\centering
\includegraphics[width=110mm]{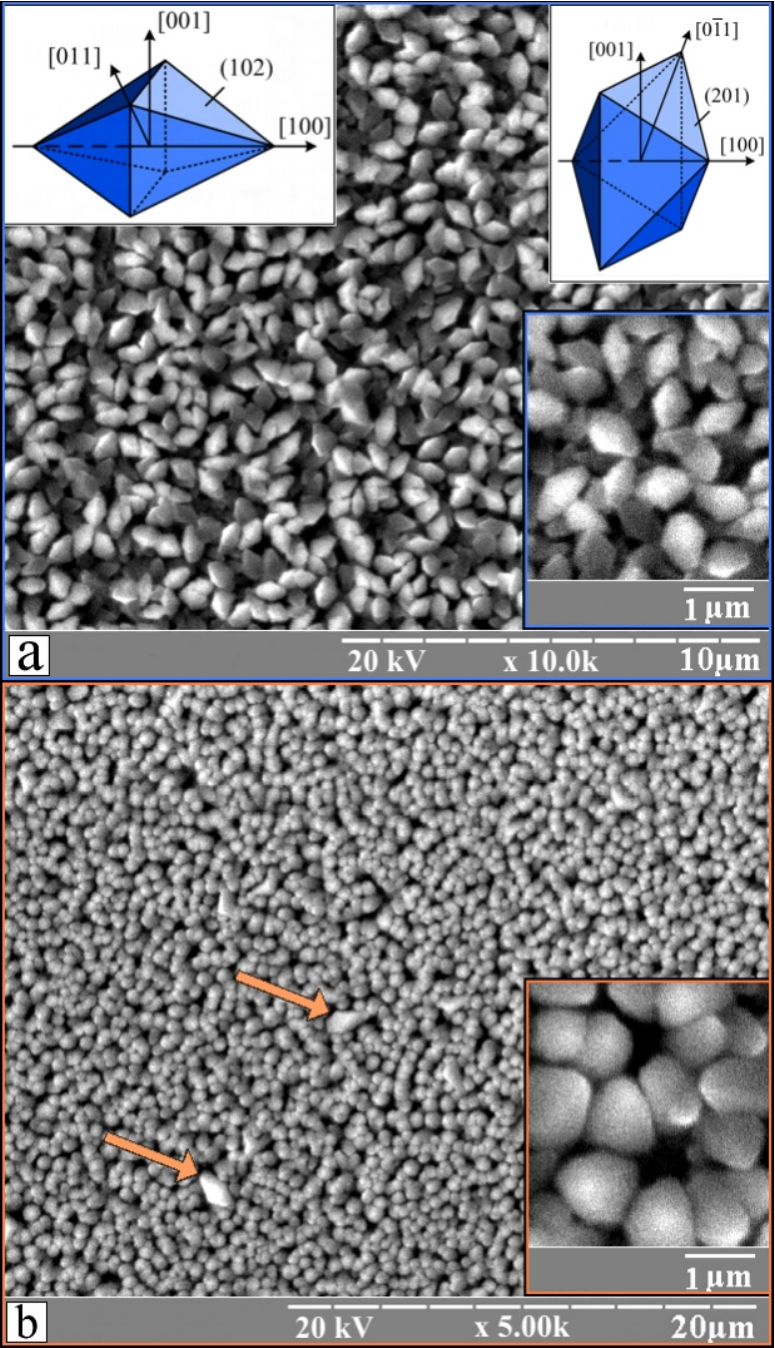}
\caption{ (Colour online) Scanning electron microscopy images of
{\it ex-situ} grown aluminium condensates at layer-by-layer (a) and
normal (b) crystal growth (the arrows on the lower panel point out
the single crystals retaining faceting of the previous structure).
There are used the glass substrates at the argon pressures $20 Pa$
(a) and $15 Pa$ (b), the discharge power $1.8 W$, the deposition
time $9 h$, the area ratio $S/s\sim 4$ (the condensate thickness is
of order $12\div14 \mu m$).}\label{Fig_3}
\end{figure}

In our experiment, we investigate the structure of the aluminium
condensates, produced in proximity to equilibrium in the
plasma-condensate system. The scanning electron microscopy images of
the ex-situ grown condensates are shown in Fig.\ref{Fig_3}. These
condensates were obtained on glass substrates by means of the
accumulative ion-plasma system described in Section \ref{Sect.2} at
the discharge power $1.8 W$ during the deposition time $t=9 h$ and
at both pressures $20 Pa$ and $15 Pa$ of the highly refined argon.
It is the argon pressure as technological parameter, that controls
supersaturation, or degree of proximity to equilibrium, at stable
discharge power. Thus, if argon pressure decreases from 20 to 15 Pa,
plasma particle concentration decreases as logical result, and,
therefore, their average energy increases. This reduces the
effective desorption energy and accordingly to Eqs. (\ref{1}) and
(\ref{2}) reduces supersaturation as well.

According to Fig.\ref{Fig_3}a, at increased pressure $20 Pa$, there
are condensed weakly bound crystals whose habitus is fixed by
crystallography planes akin to (210) because of the the conditions
$E_{(531)}<E_{c}<E_{(210)}$ inherent in energies $E_{(hkl)}$ of
atoms in half-crystal position on the $(hkl)$-planes of fcc-metals
\cite{bib:CG}. It is worthwhile to stress the structure depicted in
Fig.\ref{Fig_3}a can be statistically homogeneous up to tens of
micrometers in thickness. Such a picture of the condensate formation
is caused with the sequence of repeated nucleation on the active
centers of the growth surface being usually regions of crystallite
joining.

With reduction of the working gas pressure down to $15 Pa$, at
remaining stable of the rest of technological parameters, the
supersaturation decreases approaching the phase equilibrium of the
system plasma-condensate. This means increasing the critical energy
of chemical bonds $E_c$ which becomes higher the energy $E_{(210)}$
of atoms on the plane $(210)$, but lower the energy $E_r$ of
chemical bonds on atomically rough surface. As a result, the
conditions $E_{(210)}<E_{c}<E_r$ are fulfilled to change the
mechanism of atom attaching to growth surface. Above assumption
explains the transition from tangential to normal growth of round
shaped crystals with atomically rough surface shown in
Fig.\ref{Fig_3}. Proximity of normal and tangential mechanisms of
the crystal growth is confirmed by the fact that in Fig.\ref{Fig_3}b
several crystals labeled with arrows have faceting appropriate to
tangential growth along the (210) plane. It should be stressed as
well that, due to quasi-equilibrium conditions of the condensation
process, there is no secondary nucleation in the regions of
crystallite joining what governs formation of the columnar
structures \cite{bib:P}.

Thus, the comparison of the electron microscopy images depicted in
Figs. \ref{Fig_3}a and \ref{Fig_3}b allows for one to conclude that
the microstructures related are obtained as result of different
mechanisms of crystal growth. Obviously, above transition from
layer-by-layer to normal crystal growth may be possible at two-fold
conditions: first, the system plasma-condensate should be extremely
near the phase equilibrium (to increase the critical energy
$E_{c}$); second, condensation process should be strongly
steady-state (to fix a location of the threshold $E_{c}$ in the
chemical bond spectrum). As is shown in Section \ref{Sect.3}, both
conditions pointed out are provided due to the self-organization of
quasi-equilibrium steady-state condensation within device presented
in Section \ref{Sect.2}.

\section{Conclusion}\label{Sect.5}

In contrast to widespread technologies, we propose here original method to
realize quasi-equilibrium steady-state conditions of condensation. Character
peculiarity of this method is that condensate surface grows atom-by-atom on
active centers having the strongest chemical bonds. We show such regime is
ensured by the accumulative ion-plasma device whose advantage in comparison
with already existing systems consists in that it does not require cumbersome
control system. The self-organized regime of the device operation is confirmed
both theoretically and experimentally.

Above regime is shown to be achieved by means of two principle components of
the device to be the magnetron sputterer and the hollow cathode. Due to
diffusion, accumulation of condensed atoms occurs in the interior of the hollow
cathode up to quasi-equilibrium concentration whose value remains constant
during condensation process. Heating growth surface with the help of both
plasma stream and decreasing the effective desorption energy causes extremely
low supersaturation near condensate overall.

The physical reason to reach the quasi-equilibrium steady-state conditions is
that condensation process evolves under scenario of the
self-or\-ga\-ni\-za\-tion. On the basis of the phase-plane method, it is shown
the self-or\-ga\-ni\-za\-tion is governed by self-consistent variations of
the surface temperature and the supersaturation.

To confirm that quasi-equilibrium steady-state condensation process
is evolved under scenario of the self-organization, we investigate
the structure of the aluminium condensates obtained at different
technological conditions. Related scanning electron microscopy
images show the microstructures obtained as result of the
transforming layer-by-layer crystal growth into normal one. That, in
turn, may be possible if the system plasma-condensate is near the
phase equilibrium and condensation process is strongly steady-state.
Just that very conditions are provided due to the self-organization
of low steady-state supersaturations within device proposed.

\section{Acknowledgements}\label{Sect.6}

We are grateful to anonymous referees for constructive criticism.

\end{document}